\documentclass[a4paper]{jpconf}
\usepackage{graphicx}
\usepackage{amsfonts}
\begin{document}

\newcommand{\md}{{\rm d}}
\newcommand{\sgn}{\mathop{\rm sgn}}
\newcommand{\lP}{\ell_{\rm P}}

\mbox{}\hfill AEI-2005-022

\title{The Early Universe in Loop Quantum Cosmology}

\author{Martin Bojowald}

\address{Max-Planck-Institute for Gravitational Physics, Albert-Einstein-Institute, Am M\"uhlenberg 1, 14476 Potsdam, Germany}

\ead{mabo@aei.mpg.de}

\begin{abstract}
  Loop quantum cosmology applies techniques derived for a background
  independent quantization of general relativity to cosmological
  situations and draws conclusions for the very early universe. Direct
  implications for the singularity problem as well as phenomenology in
  the context of inflation or bouncing universes result, which will be
  reviewed here. The discussion focuses on recent new results for
  structure formation and generalizations of the methods.
\end{abstract}

\section{Introduction}

The distinguishing feature of general relativity, in comparison to
other interactions, is the fact that the metric as its basic field
does not just provide a stage for other fields but is dynamical
itself. In particular in cosmological situations the metric differs
significantly from a static background and cannot be written as a
perturbation. Thus, a faithful quantization requires a background
independent formalism which then must be non-perturbative.

An approach to quantum gravity which realizes this from the outset is
loop quantum gravity \cite{Rov:Loops,ThomasRev,ALRev,Rov}. Here,
background independence leads to a discrete structure of geometry
whose scale is a priori free (set by the Barbero--Immirzi parameter
\cite{AshVarReell,Immirzi}; in this paper we set the value equal to
one for simplicity) but fixed to be close to the Planck scale by black
hole entropy calculations \cite{ABCK:LoopEntro,IHEntro,Gamma,Gamma2}.
Thus, it is not of relevance on directly accessible scales and will
only become noticeable in high curvature regimes. In particular, this
is the case close to the big bang where the universe itself is small.

Classically, the universe would emerge from or evolve into a
singularity at those scales, where energy densities blow up and
Einstein's equations break down. For a long time, it has been hoped
that quantum gravity will resolve this problem and provide a more
complete framework which does not break down. Moreover, since this
will inevitably come with modifications of the classical theory at
small scales, one can expect phenomenological and potentially
observable consequences in the very early universe.

Even classically, it is difficult to analyze the situation in full
generality, and the quantum theory is even more complicated and less
understood. A common strategy in such a situation consists in
introducing symmetries which can be taken as homogeneity or isotropy
in the cosmological context. In contrast to earlier approaches
initiated by Wheeler and DeWitt \cite{DeWitt,QCReview}, the theory has
now been developed to such a level that the introduction of
symmetries can be done at the quantum level by employing symmetric
states \cite{SymmRed}, rather than reducing the classical theory first
and then quantizing. The relation to the full theory is thus known,
and it is possible to ensure that special features required for a
consistent background independent formulation translate to the
symmetric context.

It is then possible to take properties of the full theory, transfer
them to symmetric models and analyze them in this simpler context. In
particular, the discreteness of spatial geometry survives the
reduction \cite{cosmoII}, which is already a difference to the
Wheeler--DeWitt quantization. It also implies that there are in fact
modifications at small scales coming from the full theory, whose
phenomenological consequences can be studied in cosmological models
\cite{LoopCosRev}.

\section{Variables}

A spatially isotropic space-time has the metric
\[
 \md s^2 = -\md t^2 +\frac{a(t)^2}{(1-kr^2)^2} \md r^2+a(t)^2 r^2\md\Omega^2
\]
where $k$ can be zero or $\pm1$ and specifies the intrinsic curvature
of space, while the scale factor $a(t)$ describes the expansion or
contraction of space in time. It is subject to the Friedmann equation
\begin{equation}
 3(\dot{a}^2+k)a= 8\pi GH_{\rm matter}(a,\phi,p_{\phi})
\end{equation}
where $G$ is the gravitational constant and $H_{\rm matter}$ the
matter Hamiltonian (assumed here to be given only by a scalar $\phi$ and its
momentum $p_{\phi}$). The matter Hamiltonian depends on the matter
fields, but also on the scale factor since matter couples to
geometry. In the case of a scalar, for instance, we have
\begin{equation} \label{Hmatter}
 H_{\rm matter}=\case{1}{2}a^{-3}p_{\phi}^2+a^3 V(\phi)
\end{equation}
with the scalar potential $V(\phi)$ and the classical momentum
$p_{\phi}=a^3\dot{\phi}$.

Loop quantum gravity is based on Ashtekar variables, which provide a
canonical formulation of general relativity in terms of a densitized
triad and an SU(2) connection on space. In the isotropic context this
reduces to working with the isotropic triad component $p$ with
$|p|=a^2$ and the isotropic connection component
$c=\frac{1}{2}(k+\dot{a})$ which are canonically conjugate:
$\{c,p\}=8\pi G/3$. There is one essential difference to the metric
formulation: $p$ can take both signs since it depends on the
orientation of the triad. Thus, $p$ does not only determine the size
of space through $|p|$, but also its orientation via $\sgn p$.
(Another difference is that, when isotropic models are derived through
homogeneous ones, a canonical formalism is not available for $k=-1$.
We will thus restrict ourselves to $k=0$ and $k=1$.)

Dynamics in the canonical formulation is dictated by the Hamiltonian
constraint
\begin{equation}
  H=-3(4\pi G)^{-1}\left[2c(c-k)+k^2\right]\sqrt{|p|}+H_{\rm
matter}(p,\phi,p_{\phi})=0
\end{equation}
which indeed reduces to the Friedmann equation upon using the
definition of $p$ and $c$. Moreover, the Hamiltonian constraint $H$
gives Hamiltonian equations of motion for gravitational variables,
such as $\dot{c}=\{c,H\}$ resulting in the Raychaudhuri equation
\begin{equation}
  \frac{\ddot{a}}{a}=-\frac{4\pi G}{3a^3}\left(H_{\rm
matter}(a,\phi,p_{\phi})- a\frac{\partial H_{\rm
matter}(a,\phi,p_{\phi})}{\partial a}\right) \,,
\end{equation}
and matter equations of motion, e.g.\ for a scalar
\begin{eqnarray*}
 \dot{\phi} &=& \{\phi,H\}= p_{\phi}/a^3\\
 \dot{p}_{\phi} &=& \{p_{\phi},H\}= -a^3 V'(\phi)
\end{eqnarray*} 
which lead to the Klein--Gordon equation
\begin{equation}
 \ddot{\phi}+3\dot{a}a^{-1}\dot{\phi}+V'(\phi)=0\,.
\end{equation}

\section{Loop quantization}

While a Wheeler--DeWitt quantization would start with a Schr\"odinger
representation and work with wave functions $\psi(a)$ such that $a$ is
represented as a multiplication operator and its momentum related to
$\dot{a}$ by a derivative operator, the loop quantization implies an
inequivalent representation \cite{Bohr}. Here, one usually starts in
the connection representation such that states are functions of $c$,
an orthonormal basis of which is given by
\begin{equation} \label{states}
 \langle c|\mu\rangle = e^{i\mu c/2}, \qquad \mu\in{\mathbb R}\,.
\end{equation}
Since these states are by definition normalized, it is clear that the
Hilbert space is non-separable (it does not have a countable basis)
and that the representation is inequivalent to that assumed in the
Wheeler--DeWitt quantization.

Basic operators, which quantize $p$ and $c$, also have properties
different from $a$ as a multiplication operator or its conjugate as a
derivative operator. The action of basic operators on states
(\ref{states}) is given by
\begin{eqnarray}
\hat{p}|\mu\rangle &=&
{\textstyle\frac{1}{6}}\lP^2\mu|\mu\rangle\label{p}\\
\widehat{e^{i\mu'c/2}}|\mu\rangle &=& |\mu+\mu'\rangle \label{c}
\end{eqnarray}
with the Planck length $\lP=\sqrt{8\pi G\hbar}$. Thus, since all
eigenstates $|\mu\rangle$ of $\hat{p}$ are normalizable, $\hat{p}$ has
a discrete spectrum. Moreover, there is only an operator for the
exponential of $c$, not $c$ directly. Both properties are very
different from the corresponding operators in the Wheeler--DeWitt
quantization where the analog of $p$, the scale factor $a$, has a
continuous spectrum and its momentum has a direct quantization.

On the other hand, the properties of the basic operators (\ref{p}),
(\ref{c}) are analogous to those in the full theory, where also flux
operators quantizing the triad have discrete spectra and only
holonomies of the connection are well-defined operators but not the
connection itself. In the full theory, these properties are
consequences of the background independent formulation: One has to
smear the basic fields given by the connection and the densitized
triad in order to have a well-defined Poisson algebra to represent on
a Hilbert space. In field theory this is usually done in a
three-dimensional way using the background metric to provide a
measure. This is certainly impossible in a background independent
formulation, but there are natural, background independent smearings
of the connection along one-dimensional curves and of the densitized
triad along surfaces. Their algebra, the holonomy-flux algebra, is
well-defined and one can then look for representations. Here, it turns
out that there is a unique one carrying a unitary action of the
diffeomorphism group \cite{FluxAlg,Meas,HolFluxRep,SuperSel,WeylRep}. In
this representation, fluxes and thus spatial geometric operators which
are built from triad components have discrete spectra as a direct
consequence of background independence.

This representation is then carried over to symmetric models such that
triads have discrete spectra, too, and only exponentials of connection
components are directly represented. These properties of the loop
representation define the structure of the algebra of basic operators,
and they have far-reaching consequences:

\begin{center}
\begin{picture}(160,80)(0,0)
\put(80,70){\makebox(0,0){discrete triad \hspace{2cm} only holonomies}}
\put(30,60){\vector(0,-1){10}} \put(130,60){\vector(0,-1){10}}
\put(80,45){\makebox(0,0){finite inverse volume \hspace{2cm} discrete evolution}}
\put(40,40){\vector(1,-1){10}} \put(125,40){\vector(-1,-1){10}}
\put(80,25){\makebox(0,0){ non-singular}}
\put(30,35){\vector(0,-1){20}} \put(130,35){\vector(0,-1){20}}
\put(80,5){\makebox(0,0){\hspace{-1cm}non-perturbative modifications \hspace{1cm}
higher order terms}}
\end{picture}
\end{center}

As a consequence of the discrete triad spectrum, operators quantizing
the inverse volume are finite despite the classical divergence. This
already signals a more regular behavior at the classical singularity
which has to be confirmed by using the quantum dynamics. This
dynamics, as a consequence of the second basic quantity that only
exponentials of $c$ are represented, happens in discrete internal time.
Together with the properties of inverse volume operators this combines
to a non-singular cosmological evolution. Moreover, inverse volume
operators imply non-perturbative modifications to the classical
Friedmann equation, while the second basic property leads to
perturbative higher order terms. Both corrections have
phenomenological consequences.

\section{Non-singular evolution}

In this section we discuss the consequences of basic loop properties
as for their implication of the quantum evolution \cite{Sing,DynIn,Essay}.
In the following section we will then turn to phenomenological
consequences \cite{Inflation}.

\subsection{Finite inverse volume}

Since, as one of the basic loop effects, the triad operator $\hat{p}$
has a discrete spectrum containing zero it does not have a densely
defined inverse. On the other hand, if we want to quantize a matter
Hamiltonian such as (\ref{Hmatter}) which enters the dynamics, we
always need inverse powers of the scale factor in the kinetic term. It
seems that quantum cosmology based on a loop representation would
already come to an end at this basic step. However, there are general
methods in the full theory \cite{QSDV} which allow us to find a
well-defined quantization of $a^{-3}$. To that end we first rewrite
the classical expression in an equivalent way which is better suited
to quantization. Since such a rewriting can be done in many
classically equivalent ways, this in general leads to quantization
ambiguities in non-basic operators. For instance, the inverse volume
can be written as
\begin{equation} \label{densclass}
 d(a):=a^{-3}= \left(\frac{3}{8\pi Glj(j+1)(2
 j+1)}\sum_{I=1}^3{\rm tr}_j(\tau_I
 h_I\{h_I^{-1},|p|^l\})\right)^{3/(2-2l)}
\end{equation}
where $j\in\frac{1}{2}{\mathbb N}$ (denoting the SU(2) representation
in which we take the trace of holonomies $h_I=\exp(c\tau_I)$ with
SU(2) generators $\tau_I=-i\sigma_I/2$ in terms of Pauli matrices
$\sigma_I$) and $0<l<1$ are ambiguity parameters. The advantage of
these new expressions is that we now have only positive powers of $p$
on the right hand side which, as well as the holonomies, we can easily
quantize. The Poisson bracket will then be turned into a commutator at
the quantum level resulting in a well-defined operator whose
eigenvalues
\begin{equation}
 \widehat{d(a)}_{\mu}^{(j,l)} =
\left(\frac{9}{\ell_{\rm
P}^2lj(j+1)(2j+1)} \sum_{k=-j}^j
k|p_{\mu+2k}|^l\right)^{3/(2-2l)}
\end{equation}
on eigenstates $|\mu\rangle$ follow from the action of basic
operators.

Since this operator is finite \cite{InvScale}, the classical
divergence of $a^{-3}$ is now indeed removed as can be seen from the
eigenvalues. In particular for larger $j$, which are sometimes helpful
for phenomenological purposes, the exact expression can be difficult to
use and the approximation \cite{Ambig,ICGC}
\begin{equation} \label{deff}
  d(a)^{(j,l)}_{\rm eff}:=
 \widehat{d(a)}_{\mu(a^2)}^{(j,l)}= a^{-3} p_l(3a^2/j\ell_{\rm
   P}^2)^{3/(2-2l)}
\end{equation}
with $\mu(p)=6p/\ell_{\rm P}^2$ and
\begin{eqnarray}
 p_l(q) &=&\frac{3}{2l}q^{1-l}\left( \frac{1}{l+2}
\left((q+1)^{l+2}-|q-1|^{l+2}\right)\right.\\
 && - \left.\frac{1}{l+1}q
\left((q+1)^{l+1}-{\rm sgn}(q-1)|q-1|^{l+1}\right)\right)\nonumber
\end{eqnarray}
is helpful. Even though there are quantization ambiguities, the
important properties such as the finiteness of the operator and its
classical limit are robust.

\subsection{Difference equation}

In order to consider dynamics and to decide whether or not the
classical singularity persists as a boundary to the evolution, we need
to quantize the Friedmann equation \cite{IsoCosmo}. This is most
conveniently expressed in the triad representation given by
coefficients $\psi_{\mu}(\phi)$ in an expansion
$|\psi\rangle=\sum_{\mu}\psi_{\mu}(\phi)|\mu\rangle$ in triad
eigenstates. Since we have to use the basic operator (\ref{c}) which
is a shift operator on triad eigenstates, the quantized Friedmann
equation becomes a difference equation for $\psi_{\mu}$:
\begin{eqnarray}
&&    (V_{\mu+5}-V_{\mu+3})e^{ik}\psi_{\mu+4}(\phi)- (2+k^2)
(V_{\mu+1}-V_{\mu-1})\psi_{\mu}(\phi)\\\nonumber
&&+    (V_{\mu-3}-V_{\mu-5})e^{-ik}\psi_{\mu-4}(\phi)
  = -\frac{4}{3}\pi
G\ell_{\rm P}^2\hat{H}_{\rm matter}(\mu)\psi_{\mu}(\phi)
\end{eqnarray}
in terms of volume eigenvalues $V_{\mu}=(\ell_{\rm
  P}^2|\mu|/6)^{3/2}$. There are also possible ambiguities in this
constraint, for instance analogous to the parameter $j$ above which
have been analyzed in \cite{AmbigConstr}. Moreover, a symmetrized
version is possible, which we do not discuss here for simplicity.

The evolution dictated by this difference equation in internal time
$\mu$ does not stop at any finite value of $\mu$. In particular, we
can uniquely evolve initial values for the wave function through the
classical singularity situated at $\mu=0$. Thus, there is no
singularity where energy densities would diverge or the evolution
would stop. This comes about as a consequence of the basic loop
properties: the discreteness of spatial geometry leads to finite
operators for the inverse volume as well as evolution in discrete
internal time. Both properties enter in the demonstration of
singularity free evolution. Physically, this means that around the
classical singularity continuous space-time and with it the classical
theory dissolve. Discrete quantum geometry, on the other hand, still
makes sense and allows us to evolve to the other side of the classical
singularity.

\section{Phenomenology}

The density (\ref{densclass}) does not just give us a kinematical hint
for the removal of classical singularities, it is also important as an
ingredient in matter Hamiltonians such as (\ref{Hmatter}). Since at
small scales the classical $a^{-3}$ is modified by (\ref{deff}), we
obtain modified Hamiltonian equations of motion and a modified
Friedmann equation. For a scalar they are the effective Friedmann
equation
\begin{equation} \label{effFried}
 3a\dot{a}^2=8\pi G
\left({\textstyle\frac{1}{2}}d(a)_{\rm eff}\, p_{\phi}^2+a^3
V(\phi)\right)\,,
\end{equation}
and Raychaudhuri equation
\begin{equation} \label{effRay}
 \frac{\ddot{a}}{a}= -\frac{8\pi G}{3}\left( a^{-3}d(a)_{\rm
eff}^{-1}\dot{\phi}^2 
\left(1-{\textstyle\frac{1}{4}}a\frac{\md
\log(a^3d(a)_{\rm eff})}{\md a}\right) -V(\phi)\right)
\end{equation}
for the scale factor and the effective Klein--Gordon equation
\begin{equation} \label{effKG}
  \ddot{\phi}=\dot{\phi}\,\dot{a}\frac{\md\log d(a)_{\rm
eff}}{\md a}-a^3d(a)_{\rm eff}V'(\phi)
\end{equation}
for the scalar. Other matter types have been discussed in
\cite{Metamorph}. These modifications from $d(a)$ at small scales lead
to diverse effects which give us a new picture of the very early
universe. Before discussing these effects we note that even though
small scales behave very differently from large ones, there is an
interesting duality in the effective equations which can be helpful in
analyzing solutions \cite{JimDual}.

\subsection{Inflation}

\begin{figure}
 \begin{center}
 \includegraphics[width=12cm]{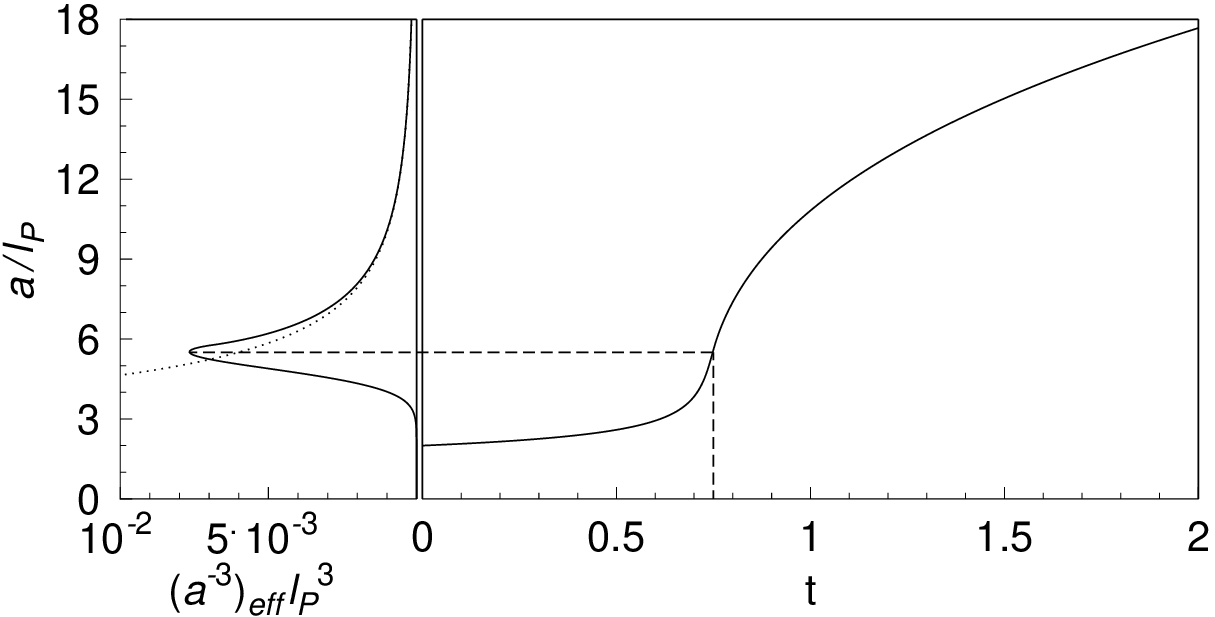}
 \caption{The effective density (left) implying early inflation
   (right), which is not realized with the classical density (dotted).
   \label{Infl}}
\end{center}
\end{figure}

At small $a$, the effective density $d(a)_{\rm eff}\sim a^{3/(1-l)}$
is increasing since $0<l<1$. Thus, in contrast to the classically
decreasing $a^{-3}$ the effective density implies a matter energy on
the right hand side of the Friedmann equation (\ref{effFried}) which
increases with the scale factor. Since the negative change of energy
with volume defines pressure, this quantum geometry effect naturally
implies an inflationary phase in early stages \cite{Inflation}. As
demonstrated in Fig.~\ref{Infl}, inflation automatically ends when the
peak of the effective density is reached such that there is no
graceful exit problem.

Since the modification is present in the kinetic term of any matter
field, we do not need to assume $\phi$ to be an inflaton with special
properties. Thus, there are several different scenarios depending on
whether or not we assume an inflaton field to drive the expansion. It
is, of course, more attractive to work without a special field, but it
also leads to complications since many of the techniques to evolve
inhomogeneities are not available. Nevertheless, recent results
\cite{GenericInfl,PowerLoop} suggest that this phase can generate a nearly
scale invariant spectrum which is consistent with observations but
also provides characteristic signatures compared to other inflation
models. However, this phase alone cannot lead to a large enough
universe (unless one assumes the parameter $j$ to be unnaturally
large) such that we need a second phase of accelerated expansion whose
properties are not restricted so tightly since it would not give rise
to observable anisotropies. 

\begin{figure}
 \begin{center}
 \includegraphics[width=12cm]{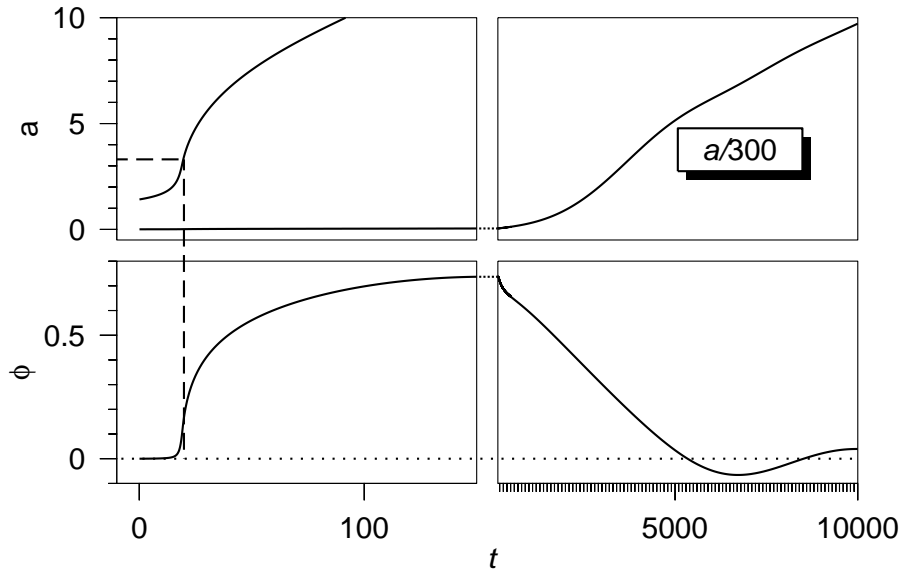}
 \caption{ During the modified phase, $a(t)$ is accelerated and $\phi$ moves
   up its potential (quadratic in this example) if it starts in a
   minimum. At the peak value of the effective density, indicated by
   the dashed lines, this first phase of inflation stops, but there
   will be a second phase (right hand side) when $\phi$ rolls down its
   potential before it oscillates around the minimum. Left hand side
   plots are in slow motion: each tic mark on the right $t$-axis
   stands at an increase of $t$ by 100. The upper right data are
   rescaled so as to fit on the same plot. Units for $a$ and $\phi$
   are Planck units, and parameters are not necessarily realistic but
   chosen for plotting purposes.
 \label{Push}}
\end{center}
\end{figure}

Such a second phase of accelerated expansion also follows naturally
from loop quantum cosmology with matter fields: The effective
Klein--Gordon equation (\ref{effKG}) has a $\dot{\phi}$-term which
changes sign at small scales. This means that the usual friction in an
expanding universe turns into antifriction very early on \cite{Closed}.
Matter fields are then driven away from their minima in the first
inflationary phase and, after this phase stops and the classical
equations become valid, slowly roll down their potentials
(Fig.~\ref{Push}). As usually, this will lead to a second (or more)
inflationary phase which makes the universe big.

This effect also applies if we do have an inflaton field. It will then
be driven to large values in the loop inflationary phase, providing
the necessary large initial values for the phase of slow-roll
inflation. Moreover, since around the turning point of the scalar
slow-roll conditions are violated, there are deviations at very early
stages of the slow-roll phase which can explain the observed loss of
power on large scales of the anisotropy spectrum
\cite{InflationWMAP,Robust}.

\subsection{Bounces}

Both effects, the modified matter energy in the Friedmann equation and
the antifriction term in the Klein--Gordon equation, can also lead to
bounces in systems which would be singular classically.  This provides
intuitive explanations for the absence of singularities
\cite{BounceClosed,BounceQualitative,GenericBounce} and can be used
for the construction of universe models \cite{Oscill,Cyclic,InflOsc}.

For a bounce we need $\dot{a}=0$, $\ddot{a}>0$ which is possible only
if we have a negative contribution to the matter energy in the
Friedmann equation. This can come from a curvature term or from a
negative potential. Classically, the second condition would then be
impossible to satisfy generically as a consequence of the singularity
theorems. In the isotropic case with a scalar, this can also be seen
from the Raychaudhuri equation (\ref{effRay}) whose right hand side is
negative in both cases: it is strictly negative with a negative
potential and, if we have a curvature term, $\dot{\phi}$ will diverge
at small $a$ and dominate over the potential. However, when the
modification becomes effective, the $\dot{\phi}^2$-term in the
Raychaudhuri equation can become positive \cite{Cyclic}. Moreover, due
to antifriction $\dot{\phi}$ will not diverge in the closed case such
that the potential can generically lead to a positive $\ddot{a}$
\cite{BounceClosed}

\subsection{Quantum degrees of freedom}

The modifications used so far only relied on non-perturbative
corrections coming from the finiteness of density operators. (In this
context also perturbative corrections from the inverse scale factor
appear above the peak, but so far they do not seem significant for
cosmology \cite{PowerPert}.) In addition, there are also perturbative
corrections which are analogous to higher order or derivative terms in
an effective action. Since the methods underlying loop quantum gravity
are canonical, deriving an effective action is not possible in a
direct manner. Nevertheless, there are methods to derive the
corresponding terms in Hamiltonian equations of motion
\cite{Perturb,Josh}, and the appearance of quantum degrees of freedom,
as would be the case for effective actions with higher derivative
terms, can also be seen here.

\begin{figure}
 \begin{center}
 \includegraphics[width=10cm]{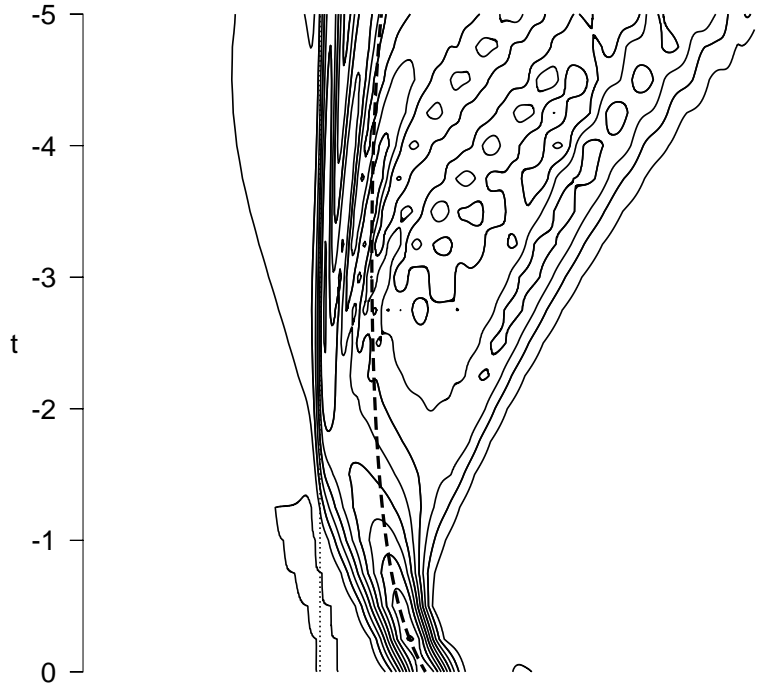}
 \caption{Contour lines of a wave packet starting at the bottom and
   moving toward the classical singularity (dotted line) where it
   bounces off, compared to a solution to the effective classical
   equation (thick dashed line) derived under the assumption of a
   Gaussian. After the bounce, the wave function departs rapidly from
   a Gaussian and deviates from the effective classical solution.
   \label{Bounce}}
\end{center}
\end{figure}

Since there are usually many different correction terms, it is not
always easy to tell which one is dominant, if any at all. This is
different from the non-perturbative effects in the density which can
be studied in isolation by choosing a large value for the ambiguity
parameter $j$. This is not always possible for other corrections, but
one can test them numerically as demonstrated in \cite{Time} where a
numerical evolution of the wave function under the Hamiltonian
constraint in coordinate time has been compared with solutions to the
effective classical equations with non-perturbative as well as
perturbative corrections. One example is a term resulting from the
spread of the wave packet which can give another explanation for
bounces (Fig.~\ref{Bounce}).

A different method to derive effective classical equations is obtained
from WKB techniques \cite{EffHam,DiscCorr}. Here, it is more difficult
to deal with new degrees of freedom arising in higher order
approximations.

\section{Less symmetric models}

All the techniques described so far for the isotropic model are now
also available for homogeneous but not necessarily isotropic models
\cite{HomCosmo,Spin}. Again, this leads to a picture at small scales
very different from the classical one. For instance, the Bianchi IX
model looses its classical chaos when modifications from effective
densities in its curvature potential are taken into account
\cite{NonChaos,ChaosLQC}. 

This also allows conclusions for the general situation of
inhomogeneous singularities where the Bianchi IX model plays an
important role in the BKL scenario \cite{BKL}. Here, space is imagined
as being composed of almost homogeneous patches each of which evolves
according to the Bianchi IX model. Thus, if the Bianchi IX model
becomes non-singular and the BKL picture remains valid, one can expect
that singularities in general are removed. However, these assumption
of a BKL picture available even at the quantum level is very strong
such that a definite conclusion can only be reached by studying
inhomogeneous midi-superspace models and eventually the full theory.
Some inhomogeneities are now under control, most importantly in
spherical symmetry \cite{SphSymm,SphSymmVol,SphSymmHam} which also
allows conclusions for black holes \cite{Horizon,Collapse}. The
singularity issue, however, is more complicated in such a situation
and remains to be resolved.

Coming back to the BKL picture, we can see that not only the structure
of classical singularities but also the approach to them is changed
dramatically in effective loop cosmology. The classical Bianchi IX
chaos implies that patches in the BKL picture have to be subdivided
rapidly if the almost homogeneous approximation is to be
maintained. This goes on without limit to the patch size which implies
unlimited fragmentation and a complicated initial geometry
classically. Without the chaos in the loop model, the fragmentation
stops eventually giving rise to a minimal patch size. As can be seen
\cite{NonChaos}, the minimal patch size is not smaller than the scale
of discreteness in loop quantum gravity thus providing a consistent
picture.

\section{Conclusions}

We have reviewed the basic ingredients of loop quantum cosmology and
discussed phenomenological consequences. Here we focused mainly on new
developments after the last report \cite{ICGC}. These are in the
context of structure formation from loop effects even without an
inflaton, new terms in effective classical equations, and better
techniques for inhomogeneous models. In particular the latter will be
developed further in the near future which will not only bring new
ingredients for cosmological investigations but also new applications
in the context of black holes.

\end{document}